\documentclass[aps,prd,showpacs,eqsecnum,nofootinbib]{revtex4}

\usepackage{epsfig}
\usepackage{graphicx}
\usepackage{dcolumn}
\usepackage{amsmath}
\usepackage{enumerate}

\def\gtap{\mathrel{ \rlap{\raise 0.511ex \hbox{$>$}}{\lower 0.511ex
   \hbox{$\sim$}}}} 
\def\ltap{\mathrel{ \rlap{\raise 0.511ex
    \hbox{$<$}}{\lower 0.511ex \hbox{$\sim$}}}}

\begin{document}

\title{
\vskip-6pt \hfill {\rm\normalsize  CERN-PH-TH/2008-037} \\[1ex]
\vskip-6pt \hfill {\rm\normalsize  IPPP/08/06} \\[1ex]
\vskip-6pt \hfill {\rm\normalsize  DCTP/08/12} \\[1ex]
\vskip-12pt~\\
MeV sterile  neutrinos in low reheating temperature 
cosmological scenarios}

\author{
\mbox{Graciela Gelmini$^{1,2}$},
\mbox{Efunwande Osoba$^{1}$},
\mbox{Sergio Palomares-Ruiz$^{3}$} and
\mbox{Silvia Pascoli$^{3}$}}

\affiliation{\vspace{2mm}
\mbox{$^1$  Department of Physics and Astronomy, UCLA,
  Los Angeles, CA 90095, USA}\\
\mbox{$^2$ CERN PH-TH, CH-1211, Geneva 23, Switzerland}\\
\mbox{$^3$ IPPP, Department of Physics, Durham University, Durham DH1 3LE, UK}
}

\vspace{6mm}

\begin{abstract}
It is commonly assumed that the cosmological and astrophysical bounds
on the mixings of sterile with active neutrinos are much more
stringent than those obtained from laboratory measurements. We point
out that in scenarios with a very low reheating temperature $T_{\rm
  RH} << 100$ MeV at the end of (the last episode of) inflation or
entropy creation, the abundance of sterile neutrinos becomes largely
suppressed with respect to that obtained within the standard
framework. Thus, in this case cosmological bounds become much less
stringent than usually assumed, allowing sterile neutrinos to be
``visible'' in future experiments. Here, we concentrate on massive
(mostly sterile) neutrinos heavier than 1~MeV.
\end{abstract}


\maketitle

\section{Introduction} 
 
In inflationary models, the beginning of the radiation dominated era of
the Universe results from the decay of coherent oscillations of a
scalar field and the subsequent thermalization of the decay products
into a thermal bath with the so called ``reheating temperature",
$T_{\rm RH}$. The standard computation of the relic densities relies
on the assumptions  that $T_{\rm RH}$ was large enough for the
particles of interest to have reached thermal equilibrium and that the
entropy of matter and radiation is conserved after they decouple. 
However, there are non-standard cosmological models
in which these assumptions about the epoch of the Universe before Big
Bang Nucleosynthesis (BBN), an epoch from which we have no data, do not
hold. These include models with moduli decay \cite{Moroi-Randall},
Q-ball decay \cite{Fujii}, and thermal inflation \cite{LS96}. In all
of these models there is a late episode of entropy production in which
the Universe is reheated to a low $T_{\rm RH}$. This temperature may
have been as low as $\sim$4~MeV~\cite{Hannestad:2004px} once all
cosmological data are taken into account, while BBN-only data impose a 
lower bound of $\sim$2~MeV on $T_{\rm RH}$~\cite{Ichikawa:2005vw} if
active neutrino oscillations are taken into account (up from
$\sim$0.7~MeV~\cite{kohri} if they are not). It is well known that a
low reheating temperature inhibits the production of particles which
are non-relativistic or decoupled at $T \ltap T_{\rm
  RH}$~\cite{giudice1,giudice2,Gelmini:2004ah}. The final number 
density of active neutrinos starts departing from the standard number
for $T_{\rm RH} \ltap 8$~MeV but stays within 10\%, 20\% or 50\% of it
for $T_{\rm RH} \gtap 5$~MeV, 4~MeV and 3~MeV, respectively. For
$T_{\rm RH} = 1$~MeV the number of tau- and muon-neutrinos would be
about 2.7\% of the standard number~\cite{giudice2,CY07}.

Low $T_{\rm RH}$ cosmological scenarios are more complicated than the
standard ones. Different aspects of these models have been
studied with interesting results, but no consistent all-encompassing
scenario exists yet. Baryogenesis could be produced through the
Affleck-Dine mechanism, in a model similar  to that of
Ref.~\cite{Dolgov:2002vf}. Dark matter could still consist of the
lightest supersymmetric particle or other Weakly Interactive Massive
Particles (WIMPs), produced either thermally or
non-thermally~\cite{GG}. Alternatively, MeV scalars have been proposed
as DM candidates~\cite{MeVDM,MeVDMdet}: they would be in thermal
equilibrium in the Early Universe with neutrinos, electrons, positrons
and photons and would freeze-out at 100's keV-- MeV temperatures.

If sterile neutrinos exist and have no extra standard-model
interactions, the dominant mechanism of production in the early
Universe is through their mixing with active
neutrinos~\cite{barbieri}. Dodelson and Widrow~\cite{dodelson} (see
also Refs.~\cite{dolgov,Abazajian:2001nj}) provided the first
analytical calculation of the production of sterile neutrinos lighter
than 100~keV in the early Universe, under the assumption (which we
maintain here) of a negligible primordial lepton asymmetry.

In general, given that the production rate of sterile neutrinos
of mass $m_s$ is maximum at a temperature $T_{\rm max} \simeq 1.3$~GeV
$(m_s/$MeV$)^{1/3} $~\cite{barbieri,dodelson}, if the reheating
temperature is smaller, $T_{\rm RH} < T_{\rm max}$, the  production of
sterile neutrinos is  suppressed. Hence, the main idea of this paper
is that the primordial abundance of sterile neutrinos does not
necessarily impose their mixing with active neutrinos to be small. A
low reheating temperature scenario would suppress the sterile neutrino
production, weakening the cosmological bounds. Thus, it might be
possible to consider massive (mostly sterile) neutrinos of any mass
and mixing with active ones, as long as laboratory bounds are
satisfied. These neutrinos could, therefore, be revealed in future
experiments. We concentrate here on the production of massive (mostly
sterile) neutrinos heavier than 1~MeV through the conversion of active
neutrinos for $T_{\rm RH} < m_s$, having already applied the same
ideas to lighter sterile neutrinos~\cite{Gelmini:2004ah}. By using
different approximations, we obtain an analytical result for the
sterile neutrino abundance. In this way, we are able to write all our
results in a simplified form. Although it is a rough approximation, it
allows us to have a qualitative understanding of the problem. For
lighter neutrinos, $m_s < 1$~MeV, the analytical calculation of
Ref.~\cite{Gelmini:2004ah} of the final sterile neutrino abundance
turned out to be correct within an order of
magnitude~\cite{CY07}. Like in Ref.~\cite{Gelmini:2004ah}, here the
active neutrinos are assumed to have the usual thermal equilibrium
distribution $f_A = (\exp{(E/T)} + 1)^{-1}$ with $E=p$. Thus,
following Ref.~\cite{Hannestad:2004px}, we restrict ourselves to
reheating temperatures $T_{\rm RH} > 4 $~MeV.

The paper is organized as follows. In Section~\ref{abundance} we
obtain an analytical approximation for the sterile neutrino
abundance as a function of the reheating temperature. In
Section~\ref{lab} we describe the laboratory bounds for sterile-active
neutrino mixing for each of the three active flavors. Using the result
of Section~\ref{abundance}, we compute in Section~\ref{cosmo} the
astrophysical and cosmological bounds and show how, in some cases, they
are completely evaded. Finally, in Section~\ref{conclusions}, we draw
our conclusions.

\section{Sterile neutrino abundance}
\label{abundance}

For simplicity, our analysis is based on the two-neutrino mixing
approximation. In this way, an analytical understanding of the problem
is possible. Within this approximation, the vacuum mixing angle $\sin
\theta$ represents the amplitude of the heavy mass eigenstate $\nu_2$
in the composition of the active neutrino flavor eigenstate
$\nu_\alpha$, i.e., $\nu_\alpha = \cos\theta~\nu_1 +\sin\theta~\nu_2
$, $\nu_s = -\sin\theta~\nu_1 +\cos\theta~\nu_2 $ for
$\alpha=e,\mu,\tau$, where $\nu_1$ is the light mass eigenstate and
$\nu_2$ is the heavy mass eigenstate, which for small $\sin \theta$ is
mostly sterile and whose mass we call $m_s$.

In order to obtain the distribution function of sterile neutrinos at a
given temperature after the last episode of inflation, we start from
the Boltzmann equation~\footnote{The distribution functions are
  defined for mass eigenstates (see the Appendix).},
\begin{equation}
\left( \frac{\partial}{\partial t} - H \, p \,
     \frac{\partial}{\partial p} \right) \, f_s = I_{\rm{coll}}~,
\label{Boltzmann-1}
\end{equation}
where $I_{\rm{coll}} \simeq \Gamma_s (T) \, (f_s^{\rm eq}-f_s)$ is the
collision integral. Here,  $f_s^{\rm eq}= (\exp{(E/T)} + 1)^{-1}$ is
the Fermi-Dirac distribution that heavy neutrinos would have if they
were in thermal equilibrium. In many cases the approximation $T \propto
1/a$, with $a$ the scale factor, is sufficiently accurate. Hence, by
using $\dot{T} = - H \, T$, Eq.~(\ref{Boltzmann-1}) can be rewritten
as (see e.g. Ref.~\cite{Dolgov1}),
\begin{equation}
-HT\left(\frac{\partial f_s} {\partial T}\right)_{p/T} \simeq
 \Gamma_s (T) \, (f_s^{\rm eq}-f_s)~.
\label{Boltzmann-2}
\end{equation}

For neutrino masses much smaller than the temperature of the plasma,
$\langle p \rangle \simeq \langle E \rangle$, ($m_s < 1$~MeV), the
averaged rate of sterile neutrino interactions is given by
\begin{equation}
\Gamma_s (T) \simeq \frac{1}{4} \, \sin^2 2 \theta_{\rm m} \,
d_\alpha \, G_{\rm F}^2 \, E \, T^4~, 
\label{GammaskeV}
\end{equation}
where $\theta_{\rm m}$ is the mixing angle in matter, $G_{\rm F}$ is
the Fermi constant and $d_{\alpha}= 1.13$ for sterile neutrino mixing
with $\nu_{\alpha}= \nu_e$ and $d_{\alpha}= 0.79$ with $\nu_{\alpha}=
\nu_{\mu,\tau}$. For $T < 1.5 \, {\rm GeV} \, (m_s/{\rm MeV})^{1/3}$
matter effects are negligible~\cite{Dolgov2} and hence $\sin^2 2
\theta_{\rm   m} \simeq \sin^2 2 \theta$ is a very good
approximation. Plugging Eq.~(\ref{GammaskeV}) into
Eq.~(\ref{Boltzmann-2}), and solving for $f_s$ in the limit $f_s <<
f_s^{\rm eq}$, the distribution function of massive (mostly sterile)
neutrinos lighter than 1~MeV was found to be~\cite{Gelmini:2004ah}
\begin{equation}
f_s (E,T)\simeq 3.2~d_{\alpha}\left(\frac{T_{\rm RH}}{5~{\rm
    MeV}}\right)^3  \sin^2{2\theta} \left(\frac{E}{T}\right) f_s^{\rm
    eq}~,
\label{old-fs}
\end{equation}
for $T_{RH} \ll T_{\rm max}$. This distribution results in a number
density of light ($m_s < 1$~MeV) sterile neutrinos given by
\begin{equation}
n_s \simeq 10 \, d_{\alpha} \, \sin^2 {2\theta}
\left(\frac{T_{\rm RH}}{5~{\rm MeV}}\right)^3 \, n_\alpha ~, 
\end{equation}
where $n_\alpha = 0.09 \, g \, T^3$ is the number density of a
relativistic fermion with $g$  degrees of freedom in thermal
equilibrium. Notice that the  number density of sterile neutrinos
depends on both the active-sterile mixing angle and the reheating
temperature. A low reheating temperature implies a small sterile
number density, even for active-sterile mixing angles as large as
experimental bounds permit (see below). However, Eq.~(\ref{old-fs}) 
is only valid if the condition $f_s << f_s^{\rm eq}$ is satisfied. We
would like to extend here this result, for the case when this
approximation starts to fail. This was not included in
Ref.~\cite{Gelmini:2004ah}. Nevertheless, Eq.~(\ref{Boltzmann-2}) can
be solved perturbatively and Eq.~(\ref{old-fs}) should be replaced by 
\begin{equation}
f_s (E,T)\simeq \left(1- e^{-S}\right) f_s^{\rm eq} (E,T)~,
\end{equation}
where $S= 3.2~d_{\alpha}\left({T_{R}}/{5~{\rm MeV}}\right)^3
\left({E}/{T}\right) \, \sin^22\theta$ is the coefficient multiplying
the equilibrium neutrino distribution in Eq.~(\ref{old-fs}). The
sterile neutrino number density $n_s$ results from a numerical
integration of this distribution. This perturbative solution is valid
for $S<1$.

For heavier sterile neutrinos, with $m_s > 1$~MeV, and for the range
of temperatures explored here, the heavy neutrino mass needs to be
taken into account and the averaged production rate of sterile
neutrinos $\Gamma_s$ is given by~\cite{Dolgov1,Dolgov2}
\begin{equation}
\Gamma_s (T) = \frac{1}{\tau_s}\left[\frac{m_s}{E} +
\frac{3 \times 2^7 \, T^3}{m_s^3}\left\{\frac{3 \, \zeta(3)}{4} +
\frac{7 \, \pi^4}{144}\left(\frac{E T}{m_s^2} + \frac{p^2T}
{3 \, E \, m_s^2}\right)\right\}\right]~,
\label{Gamma-s}
\end{equation}
where the first term is due to inverse decay and the other terms
correspond to two-to-two particle processes. The last term in
parenthesis, $\sim T^4/ \tau_s m_s^5$, due to oscillations, is the
only one remaining as $m_s\to 0$~\footnote{We note that in this limit
  there is a factor of 2 with respect to Eq.~(\ref{GammaskeV}) already
  present in the previous literature, which does not change our
  conclusions and we do not attempt to correct here.}. The function 
$\tau_s$ in the denominator is the heavy neutrino lifetime. As in the
case of $m_s < 1$~MeV, Eq.~(\ref{Gamma-s}) is valid when matter
effects are not important.

For $m_s < m_\pi \sim 140$ ~MeV, the massive (mostly sterile) neutrino
can decay into a light neutrino and two leptons $\nu_s \rightarrow
\nu_\alpha +l +\bar l $, mainly  $\nu_\alpha\bar\nu\nu$ and
$\nu_\alpha e^+e^-$. If the active neutrino mixing with the sterile 
is $\nu_{\tau}$ or $\nu_{\mu}$, the decay of $\nu_2$ happens through
neutral currents and the lifetime is
\begin{equation}
\tau_s = \frac{1.0~{\rm sec}}{\sin^22\theta}
\left(\frac{10~{\rm MeV}}{m_s}\right)^5~.
\label{life1}
\end{equation}
If instead $\nu_s$ mixes mostly with $\nu_e$, the factor  1.0 sec
should be replaced by 0.7 sec~\cite{Dolgov2, Dolgov1}, due to the
presence of charged currents. However, we are not going to keep this
distinction in the following. For $m_\pi<m_2<2m_\mu$, $\nu_2$ decays
mostly into $\pi^0\nu$, $\pi^+ e^-$ and $\pi^- e^+$, and the decay is
much faster than Eq.~(\ref{life1})~\cite{Hansen:2003yj,Dolgov2}. For
even larger masses, other decay modes open up. In the following we
will restrict ourselves to the range $m_s < m_\pi \simeq 140$~MeV,
which is enough for our purposes of showing the main characteristics
of the low reheating temperature cosmological scenarios we envision. 

Assuming that the bulk of sterile neutrinos are produced after the
reheating of the Universe~\footnote{It was shown in Ref.~\cite{CY07}
  that for masses lighter than 1~MeV this gives results which are
  correct within an order of magnitude~\cite{Gelmini:2004ah}.}, namely
that $f_s \simeq 0$ at $T= T_{\rm RH}$, we solve analytically
Eq.~(\ref{Boltzmann-2}) for $T_{\rm RH} \leq m_s$, after plugging
Eqs.~(\ref{Gamma-s}) and (\ref{life1}) into it. In order to
analytically solve the equation, we will make several
approximations. First, we assume that the actual distribution function
of the heavy neutrinos is always much smaller than the equilibrium
distribution, $f_s << f_s^{\rm eq}$. Then, we approximate the
Fermi-Dirac distribution by a Boltzmann distribution, $f_s^{\rm eq}
\simeq e^{-E/T}$ and take neutrinos to be either purely
non-relativistic, i.e., $E_2 = m_s$ if $p_2 < m_s$, or purely
relativistic, i.e., $E_2 = p_2$ if $p_2 > m_s$. We define $y \equiv 
p/T$, and integrate analytically Eq.~(\ref{Boltzmann-2}) (with $y$
constant) over temperatures $T$ in the interval $0\leq T \leq
T_{\rm RH}$. We find $f_I$ and $f_{II}$ given by 
\begin{equation}
 f_I \left(T, y\right)= \int_{{m_s}/{y}}^{T_{\rm RH}}
 \frac{\Gamma_s\left(T\right)}{H \, T} \, e^{-y} \, dT + \int_0^{m_s/y}
 \frac{\Gamma_s\left(T\right)}{H \, T} \, e^{-m_s/T} \, dT~,
\end{equation}
for $0 \leq m_s / y \leq T_{\rm RH}$ and
\begin{equation}
f_{II}\left(T, y\right)=
\int_0^{T_{\rm RH}} \frac{\Gamma_s\left(T\right)}{HT}e^{-m_s/T} \, dT~,
\end{equation} 
for $ T\leq T_{\rm RH}\leq {m_s}/{y}$. Notice that due to the Boltzmann
factor the contribution of the integrands at low temperatures is
negligible. Thus, cutting the integrations at the decoupling
temperature of active neutrinos or extending them to zero temperatures 
does not change the integrals in any significant way.

\begin{figure}
\centerline{{\epsfxsize=4.5in \epsfbox{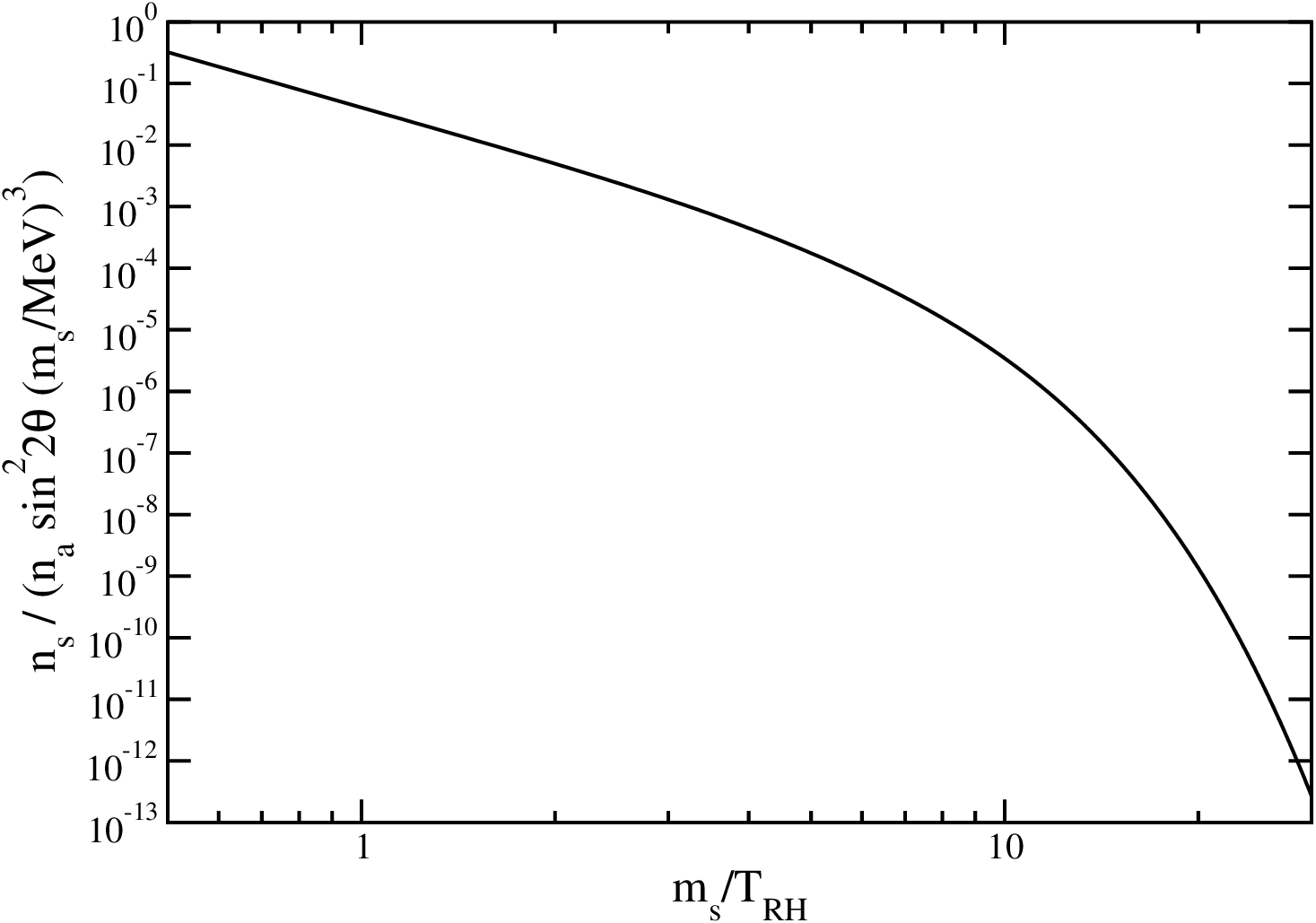}}}
\caption{
Number density of (mostly) sterile neutrinos for 1~MeV $< m_s <
140$~MeV  and 5~MeV $\leq T_{\rm RH} \leq m_s$.}  
\label{ns}
\end{figure}

Once we have the distribution function, we then find  the sterile
neutrino number density as function of the ratio ${m_s}/{T_{\rm RH}}$
and the temperature $T$, by integrating $f_I$ and $f_{II}$,
\begin{equation}
n_s(m_s/T_{\rm RH},T) =
\int_0^{T({m_s}/{T_{\rm RH}})} \frac{p^2 \, dp}{2 \, \pi^2} \,
f_{II}\left(p, T\right) 
+ \int_{T({m_s}/{T_{\rm RH}})}^{\infty} \frac{p^2 \, dp}{2 \, \pi^2} \,
f_{I}\left(p,T\right)~.   
\end{equation} 
This procedure overestimates the final abundance, thus providing an
upper bound on the actual abundance. In this way we find that
the number density $n_s$ of heavy (mostly sterile) neutrinos for $ m_s
< 140$~MeV is given by 
\begin{eqnarray}
n_s(x_{\rm RH},T) & \simeq & n_\alpha (T) \, \sin^22\theta \,
      \left(\frac{m_s}{{\rm MeV}} \right)^3 \, 2.1 \times 10^{-3} \,
      e^{-x_{\rm RH}} \nonumber \\  
& & \times  \biggl[ \left(\frac{7}{3} + \frac{6 + 144 \,
      \zeta(3)}{\pi^4}\right) +  
    \frac{2^3 \times 7}{3} \, \left\{1 + x_{\rm RH} +
      \left(\frac{3}{2^3}+\frac{3^4 \, \zeta(3)}{7 \, \pi^4}\right)
      x_{\rm RH}^2 \right\} \, \frac{1}{x_{\rm RH}^3}        
      \nonumber \\ 
& + & \left. \left(24 + 144 \, \zeta(3) + 
    12 \, x_{\rm RH} + \frac{7}{2} \, x_{\rm RH}^2 + \frac{1}{2} \,
      x_{\rm RH}^3 \right) \frac{x_{\rm RH}}{4 \, \pi^4} \right]~.
\label{number-density}
\end{eqnarray}
Here we have defined $x_{\rm RH} \equiv m_s/T_{\rm RH}$. This number
density is plotted in Fig.~\ref{ns}, where $n_s/ [n_\alpha \,
\sin^22\theta \, (m_s/\rm{MeV})^3]$ is shown as function of $x_{\rm
RH}$. Taking into account the subsequent decay of sterile neutrinos,
the actual number density is $n_s (T) \simeq  n_s (x_{\rm RH},T) \,
e^{-{\rm t}/\tau_s}$. In order to obtain analytical results for the
cosmologically and astrophysically allowed regions in the parameter
space $(\sin^2 {2 \theta}, m_s)$, in Section~\ref{cosmo} we will use
the instant-decay approximation for $\tau_s$ smaller than the age of
the Universe.

We consider next the experimental bounds and then the cosmological
and astrophysical bounds on the mass and mixing angle of heavy (mostly
sterile) neutrinos. We will show that if the reheating temperature
turns out to be sufficiently smaller than the neutrino mass, the
cosmological bounds become irrelevant and the mixing angles for any
given mass can be as large as the experimental bounds permit, making
the detection at the reach of future laboratory experiments.

\section{Experimental Bounds}
\label{lab}

\begin{figure}[t]
\includegraphics[width=0.8\textwidth,clip=true,angle=0]{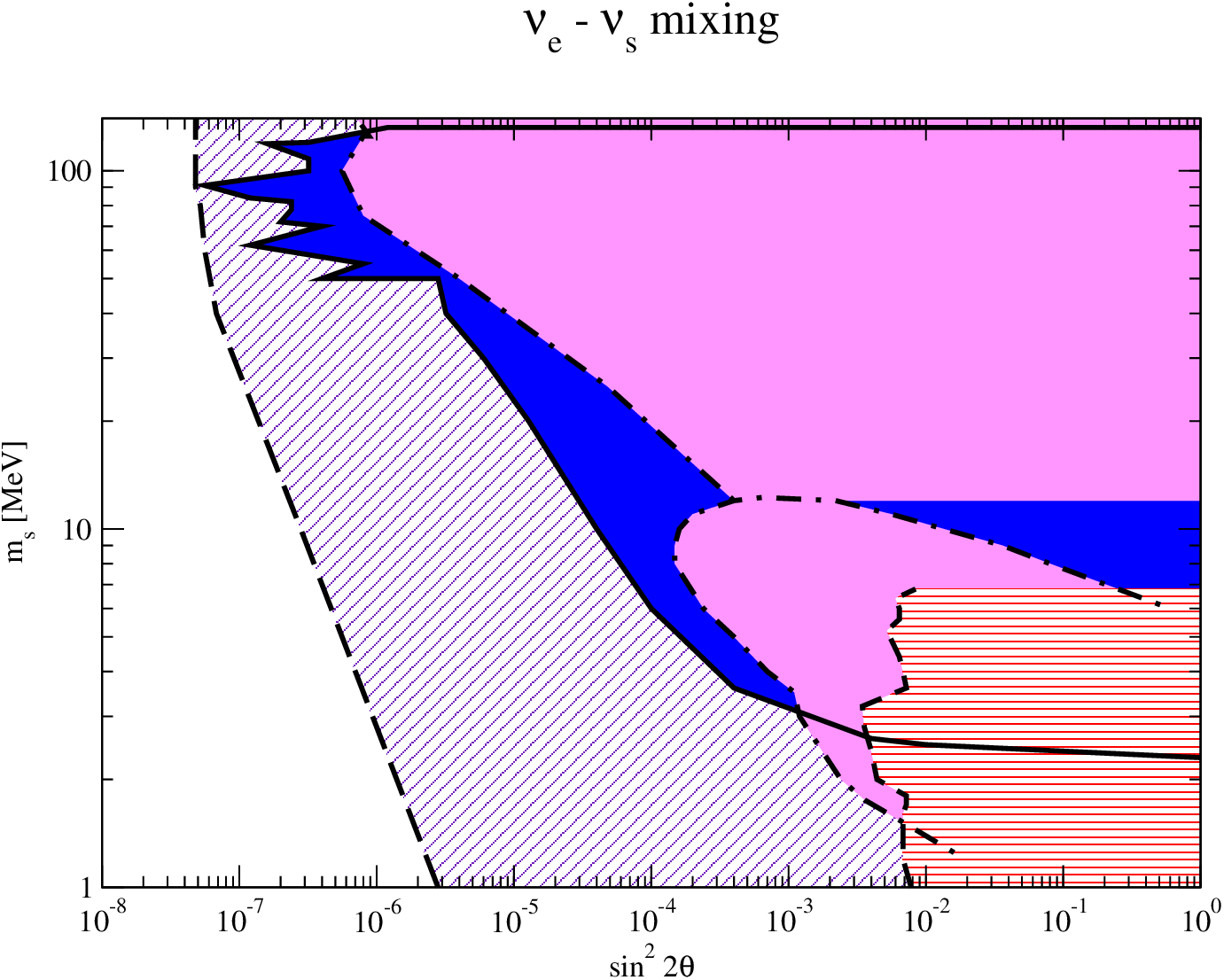} 
\caption{
Experimental bounds on the mass and mixing angle of sterile neutrinos
mixed with $\nu_e$ in the mass range
$1~{\rm{MeV}}<m_s<140~{\rm{MeV}}$. The colored regions are excluded 
by searches of: i) kinks in $\beta$-decays: horizontally hatched (red)
area with short-dashed boundary; ii) sterile neutrino decays in
visible particles: solid light gray (magenta) regions with dash-dotted
contours; iii) peaks in the electron spectrum in pion and kaon decays:
solid dark gray (blue) region delimited by a solid line; iv)
neutrinoless double beta-decay: diagonally hatched (blue) area with
long-dashed contour. See text for details.}  
\label{nues}
\end{figure}

\begin{figure}[t]
\includegraphics[width=0.8\textwidth,clip=true,angle=0]{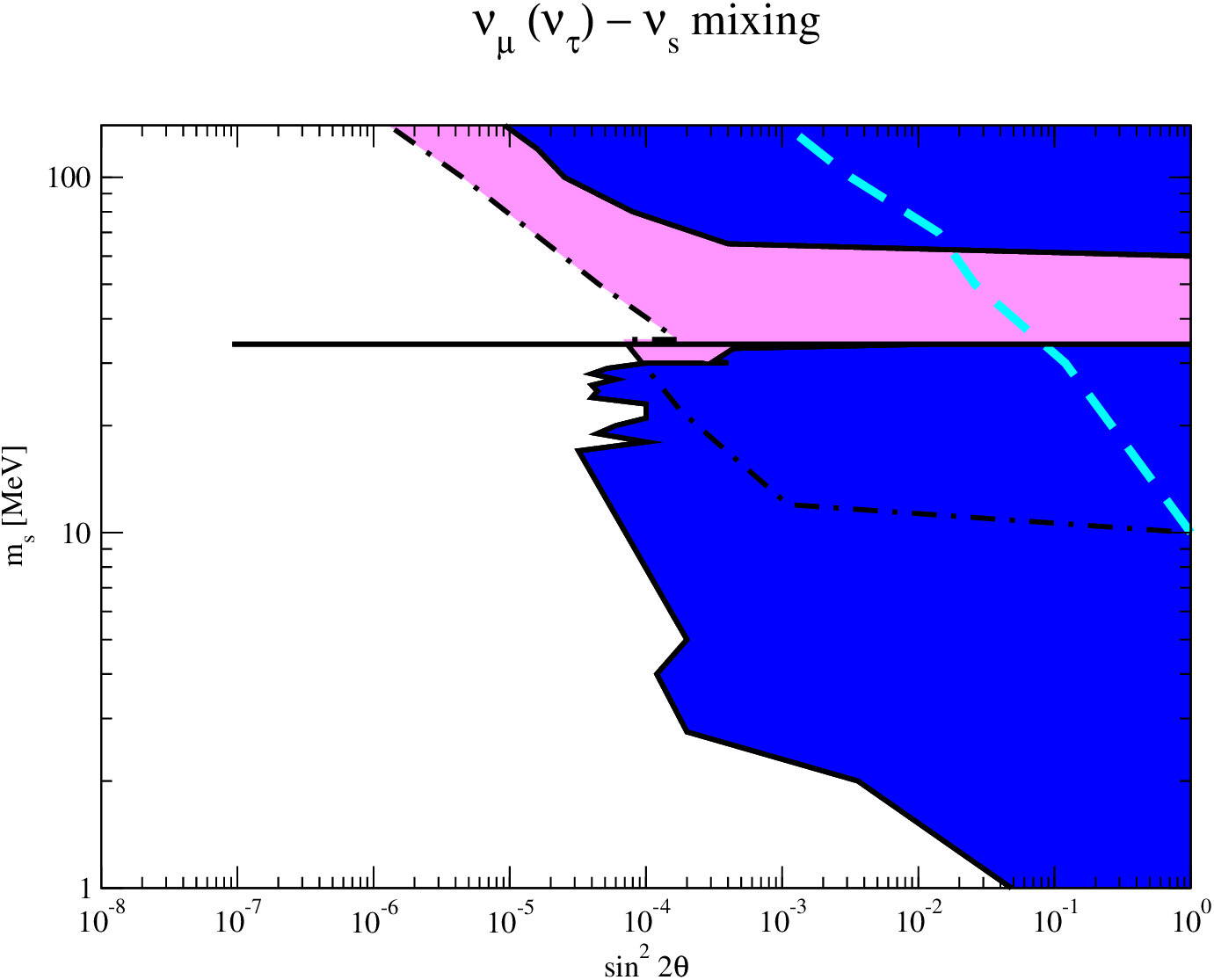}
\caption{
Experimental bounds on the mass and mixing angle of sterile neutrinos
mixed with $\nu_\mu$ (or with $\nu_\tau$) in the mass range
$1~\mathrm{MeV}<m_s<140~\mathrm{MeV}$. For the $\nu_\mu$ case, the
colored regions denote the bounds obtained from i) peak searches in
the muon spectrum from pion and kaon decays which exclude the dark gray
(blue) areas with solid contours; ii) searches of sterile neutrino
decays excluding the light gray (light-magenta) region delimited by a
dash-dotted line. The (cyan) long-dashed line represents the upper
bound for sterile neutrinos mixed with $\nu_\tau$. See text for
further details.}
\label{numus}
\end{figure}

In laboratory searches, no positive evidence of heavy (mostly sterile)
neutrinos has been found so far in the mass range of interest,
1~MeV--140~MeV. Here, we review the most stringent bounds on the mixing
angle with active neutrinos and show them in Figs.~\ref{nues} and
\ref{numus} (for further details see a comprehensive discussion in
Ref.~\cite{pascolihan}).

Let us consider first sterile neutrinos mixing with $\nu_e$. For
masses up to $m_s\simeq$~10~MeV, an important bound is provided by
searches of kinks in the electron spectrum of $\beta$-decays, which
constrain the mixing angle to be $\sin^2 {2 \theta} \ltap 6 \times
10^{-3}$, as shown in the horizontally hatched (red) excluded region in
Fig.~\ref{nues}. 

For higher masses, very robust bounds can be set by looking for
additional peaks in the spectrum of electrons in leptonic decays of
pions and kaons. The electron energy of the possible monochromatic
line, given by $E_l= (m_{\pi, K} + m_e^2 - m_s^2)/ 2 m_{\pi, K}$,
depends on the mass of the heavy sterile neutrino, while the
branching ratio is proportional to $\sin^2 {2 \theta}$. Here, $m_{\pi,
  K}$ is the mass of either the pion or the kaon respectively, $m_e$
is the mass of the electron and $m_s$ is the sterile neutrino mass. At
present, bounds as stringent as $\sin^2 {2 \theta} <10^{-7}$ are
obtained in this way (for a review see Ref.~\cite{Britton:1992pg})
which are shown in Fig.~\ref{nues} as the solid dark gray (blue)
excluded area.

In neutrino-oscillation, fixed-target and collider experiments, if
sterile neutrinos mix with active ones, a beam of $\nu_s$ would be
produced and would subsequently decay into visible particles.
Assuming only charged current and neutral current interactions for
$\nu_s$, the absence of $\nu_s$-decay signatures in past and present
experiments allows one to put limits on the mixing term which controls
the intensity of the $\nu_s$ beam and the decay time. A reanalysis of
the Borexino Counting Test Facility and Bugey data yields $\sin^2 {2
\theta}  < 10^{-4}$ for $m_s < 10$~MeV~\cite{Back:2003ae} at
90\% confidence level (CL), while the data from the experiment
PS191~\cite{Bernardi:1987ek} puts a bound which is strongly mass 
dependent, going from $\sin^2 {2 \theta} <  4 \times 10^{-4}$ at 
$m_s \sim 12$~MeV to $\sin^2 {2 \theta}  < 10^{-8}$ at $m_s \sim
300$~MeV. These constraints exclude the solid light gray
(light-magenta) regions in Fig.~\ref{nues}.

Finally, if sterile neutrinos are Majorana particles, they would
contribute to the mediation of neutrinoless double-beta decay. The
limit on the half-life time of this process can be translated into a
bound on the mixing with $\nu_e$, $\sin^2 \theta$, which scales as
$m_s$ for $m_s \ltap 30$~MeV and as $m_s^{-1}$ for $m_s \gtap
400$~MeV, with $\sin^2 {2 \theta} < 5 \times 10^{-8}$ at $m_s\sim
100$~MeV. This limit excludes the diagonally hatched (blue) region
shown in Fig.~\ref{nues}.

Let us now consider the case of mixing with $\nu_\mu$. As discussed
above for the case of sterile neutrinos mixing with $\nu_e$, peak
searches provide very robust and stringent bounds on sterile
neutrinos. Looking this time for peaks in the spectrum of muons in 
pion decays, it is possible to constrain the relevant mixing $\sin^2
{2 \theta}$ to be typically $< 10^{-4}$ in the mass range 4--34~MeV
(for a detailed review see Fig.~1 of Ref.~\cite{Kusenko:2004qc}).
Motivated by the KARMEN anomaly, which could be explained by the
existence of a heavy (mostly sterile) neutrino with mass 33.9~MeV,
very sensitive searches have been performed for this neutrino mass and
the bound was found $\sin^2 {2 \theta} < 9.2 \times 10^{-8}$ at
95\%~CL~\cite{prodpion7} (see the black horizontal line in
Fig.~\ref{numus}). Similarly, sterile neutrinos with heavier masses can
be probed in kaon decays and, in the mass range of interest, the
derived bounds go from $\sin^2 {2 \theta} < 4 \times 10^{-4}$ at
$m_s=65$~MeV to $\sin^2 {2 \theta} < 25.2 ~ (4) \times 10^{-6}$ for
$m_s= 100~\mbox{MeV} ~(200 ~\mbox{MeV})$ at 90\%~CL~\cite{Hayano82}.
These peak-searches exclude the dark gray (blue) areas shown in
Fig.~\ref{numus}.
 
By searching for the production of $\nu_s$ in pion and kaon decays and
their subsequent decay, a reanalysis~\cite{Kusenko:2004qc} of
fixed-target data led to a stringent mass-dependent bound on $\sin^2
{2 \theta}$, which exclude the light gray (light-magenta) area in
Fig.~\ref{numus}. The bound is $\sin^2 {2 \theta} < 10^{-4}$ for
masses up to 35~MeV and reaches $\sin^2 {2 \theta}< 3.4 \times
10^{-7}$ at $m_s = 200$~MeV, at 90\%~CL. Similar bounds can be set on
the product of the mixing angle with $\nu_\mu$ and with $\nu_e$ from
the analysis of the same decay-searches data (a review is given in
Ref.~\cite{Kusenko:2004qc}).

Let us consider finally the case of heavy (mostly sterile) neutrinos
mixed with $\nu_\tau$. The only limits on these sterile neutrinos come
from searches for $\nu_s$ decays. The most stringent bound is obtained
by the reanalysis of data from the CHARM experiment, in which $\nu_s$
could be produced in $D$ and $\tau$ decays. However, in the mass range
we are considering, this upper bound, shown in Fig.~\ref{numus} by the
(red) long-dashed line, is rather weak, such as $\sin^2 {2 \theta} < 2
\times 10^{-3}$ for $m_s = 100$~MeV.

\section{Cosmological Bounds}
\label{cosmo}

In what follows we obtain four different types of cosmological and
astrophysical bounds on the parameter space of active-sterile neutrino
mixing, $(\sin^2 2\theta, m_s)$. Each of these limits is valid for a
certain range of values of the heavy neutrino lifetime, $\tau_s$.
Whereas the first three bounds we describe have to do with the photons
which are produced in the decay, the last one represents a bound on
the sterile neutrino abundance at a particular epoch. Finally, we will
also comment on bounds from core collapse supernovae observations. 

In the first place, we consider the diffuse extragalactic background
radiation (DEBRA) spectrum and set bounds on the basis of not finding
any unexpected result from heavy (mostly sterile) neutrino
decay. Secondly, we obtain bounds using the non-observation of a
distortion of the Cosmic Microwave Background (CMB) spectrum caused by
the radiation from decay. Then, we calculate the limits based on the
data from the primordial light element abundances and how neutrino
decay would affect them. Finally, we also use the upper bound from
BBN on the extra number of relativistic degrees of freedom at that
epoch to set a limit on active-sterile neutrino mixing. In our
analysis, for $\tau_s < t_U$ with $t_U \simeq 14$~Gyr the present age
of the Universe, we use the instant-decay approximation, i.e., all
decays happen at $t = \tau_s$. This will allow us to obtain all the
results analytically, while giving rise to accurate enough
calculations for our purposes. 

For sterile neutrino masses in the range of interest, 1~MeV $< m_s <
140$ MeV, the decay branching ratio of $\nu_2$ into $e^+ e^-$ is about
10\% (40\%) for mixing with $\nu_\mu$ or $\nu_\tau$ ($\nu_e$). Due to
inverse Compton scattering on the CMB with a Thompson cross section,
the interaction length of these electrons is about 1~kpc (see for
example Ref.~\cite{Armengaud:2006fx} or Fig.~5 of
Ref.~\cite{Protheroe:1995ft}) and the result of each of the
interactions is a photon that shares a large portion of the incoming 
electron energy. Photons at these energies propagate for cosmological
distances undisturbed by the CMB or infrared backgrounds and are
subjected to different cosmological bounds, which depend on the time
at which the photons were produced.
 
Photons decouple from the plasma filling the Universe at the
recombination epoch, $t_{\rm rec} \simeq 1.3 \times10^{13}~{\rm
  sec}$. Thus,  if the sterile neutrino decays  happen after
recombination, $\tau_s>t_{\rm rec}$, the photons produced do not
interact ever after and could leave an imprint in the DEBRA
spectrum~\cite{EGRET}. Such a signature has not been observed, and
thus the photon flux must not be larger than the observed DEBRA energy
flux $I_\gamma \equiv E_\gamma^2 dF_\gamma/dE_\gamma$. From the
observations by the COMPTEL~\cite{COMPTEL} instrument we obtain
approximately
\begin{equation} 
I_\gamma \simeq 0.01  \, \frac{\rm{MeV}}{\rm{cm}^2 \, \rm{sec} \,
  \rm{sr}}~, 
\end{equation}
for 1~MeV $< f_\gamma m_s< $ 30~MeV, where $f_\gamma$ is the average
fraction of the sterile neutrino mass that goes into photons in each
decay. Thus $f_\gamma m_s$ is the average energy going into photons
per decay. From EGRET data~\cite{EGRET} we obtain 
\begin{equation}
I_\gamma \simeq 2.0 \times 10^{-3} \, \frac{\rm{MeV}}{\rm{cm}^2 \,
  \rm{sec} \, \rm{sr}}~, 
\end{equation}
for 30~MeV $ < f_\gamma m_s <$ 140~MeV. 
\begin{figure}[t]
\includegraphics[width=0.8\textwidth,clip=true,angle=0]{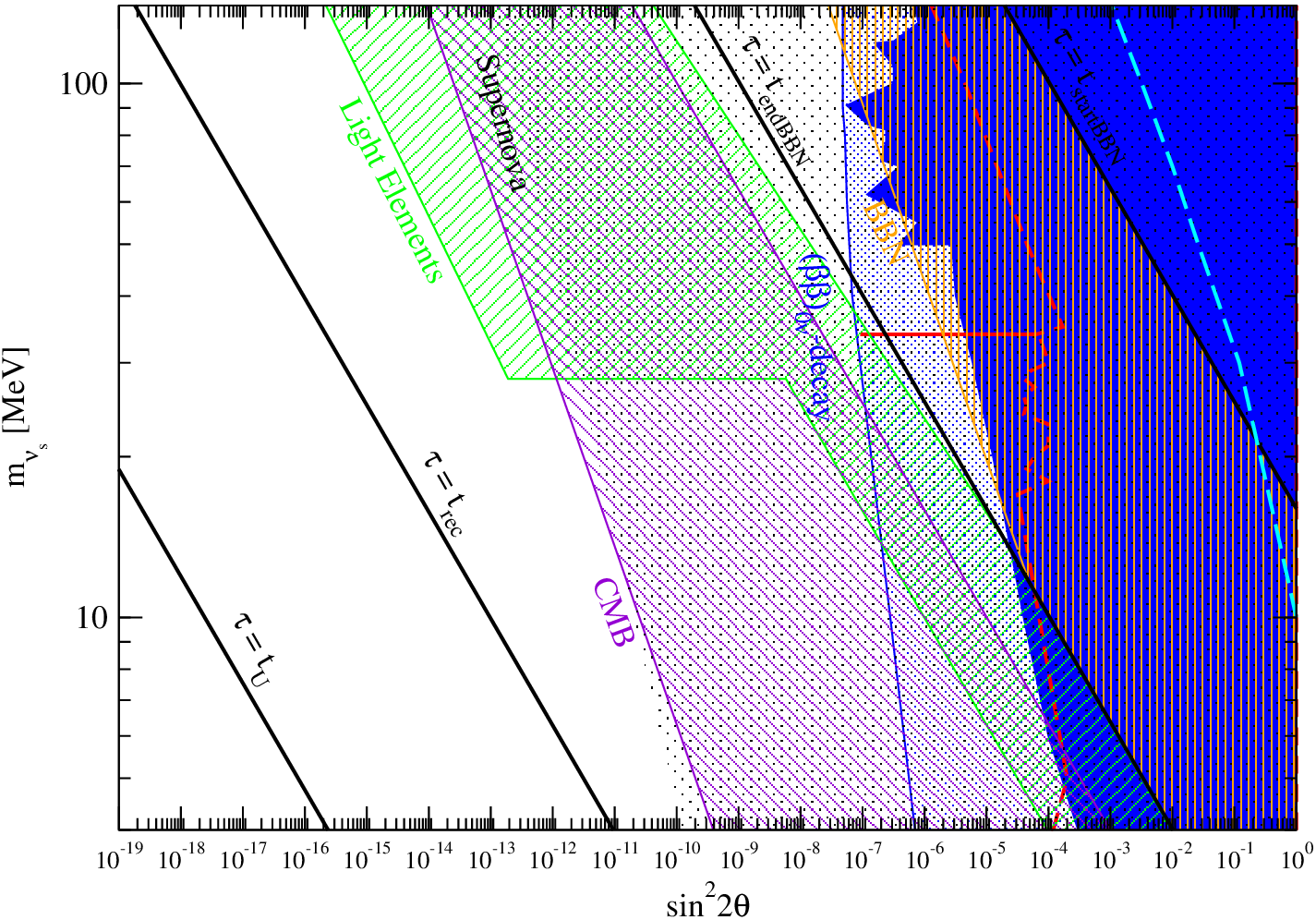}
\caption{Experimental and cosmological bounds as function of the
  mixing angle and mass of heavy (mostly sterile) neutrinos mixed with
  active ones for $T_{\rm RH} = m_s$. The dark gray (blue) solid area
  and dark gray (blue) hatched area (for Majorana neutrinos) represent
  the experimentally excluded region for $\nu_e-\nu_s$ mixing, the
  (red) short-dashed line and the (cyan) long-dashed line are the
  experimental upper bounds for $\nu_\mu-\nu_s$ and $\nu_\tau-\nu_s$
  mixing, respectively. The cosmological and astrophysical bounds are
  indicated by the corresponding labels. Isolines for values of the
  heavy neutrino lifetime $\tau$ equal to the start and end of the BBN
  epoch, the recombination time and the present age of the Universe
  are also shown. Only values of the reheating temperature larger than
  4~MeV are considered. See text for details.} 
\label{ratio-1}
\end{figure}
In the instant-decay approximation, for $\tau_s < t_U$, the energy of
the photons produced in the decays at $t=\tau_s$ redshifts from the
time $t = \tau_s$ until today. For $\tau_s > t_U$, because the decay
rate increases with time for $t < \tau_s$, most of the decays are
happening at present, and we assume there is no significant redshift
of the initial photon energy.
For $\tau_s > t_{\rm U}$, we get  
\begin{equation}
f_\gamma \, B \, m_s \, \frac{3}{2} \, \left(\frac{t_{\rm
    U}}{\tau_s}\right) \, n_s \, \frac{c}{4\pi} < I_\gamma
\end{equation}
and for  $t_{\rm rec} < \tau_s < t_{\rm U}$,
\begin{equation}
f_\gamma \, B \, m_s \, \left(\frac{\tau_s}{t_{\rm U}}\right)^{2/3} \, n_s \, 
\frac{c}{4\pi} < I_\gamma~,
\end{equation}
where $B$ is the branching ratio of the sterile neutrino decay into
photons or charged particles. We label these bounds as ``DEBRA'' and
it only appears in Fig.~\ref{T=5}. 

On the other hand, the CMB radiation is emitted at
recombination. Electromagnetic decay products produced sometime before
recombination may distort the CMB
spectrum~\cite{Ellis:1990nb,Hu:1993gc}. Non-thermal photons produced
before the thermalization time $t_{\rm th} \simeq 10^6~{\rm sec}$ are
rapidly incorporated into the Planck spectrum. This happens through
processes that  change the number of photons, such as double Compton
scattering ($\gamma e \to \gamma \gamma e$). If non-thermal photons
are produced after $t_{\rm th}$, i.e., if  $t_{\rm th}  < \tau_s <
t_{\rm rec} $, the CMB Plank spectrum would be distorted. Current data
pose very stringent upper bounds on possible distortions of this
spectrum.
For the earliest part of this last time interval, i.e., for $t_{\rm th}
\simeq 10^6~{\rm sec} < \tau_s < 10^9~{\rm sec}$, photon number
preserving processes, like elastic Compton scattering, are still
efficient. These processes thermalize the photons not into a Plank
spectrum but into a Bose- Einstein spectrum with a non-zero chemical
potential $\mu$. If the initial spectrum has fewer photons than a
black body of the same total energy, then the chemical potential is
positive $\mu>0$ (if it has more photons, then $\mu<0$). So, for
$t_{\rm th} \simeq 10^6~{\rm sec} < \tau_s < 10^9~{\rm sec}$, the
energy released into photons in the decay for $|\mu| <<1$ (the only
values of $\mu$ allowed by observations) is 
\begin{equation}
\frac{\Delta\rho_\gamma}{\rho_\gamma} \simeq  0.714 \mu~.
\label{Deltarho1}
\end{equation}
The bound provided by the COBE satellite is $|\mu| < 0.9 \times
10^{-4}$ at the 95\% CL~\cite{Fixsen:1996nj}. 
\begin{figure}[t]
\includegraphics[width=0.8\textwidth,clip=true,angle=0]{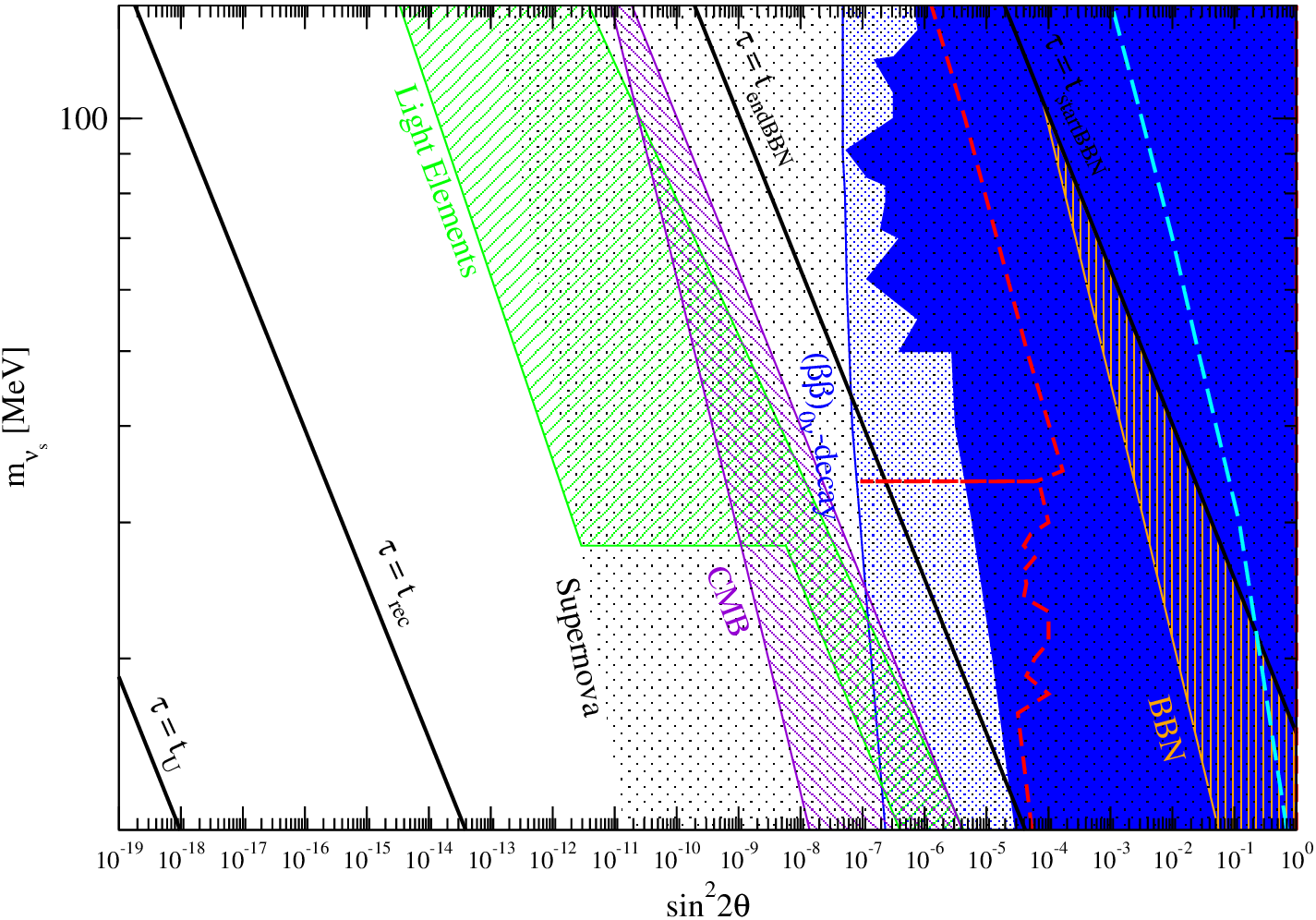}   
\caption{Same as Fig.~\ref{ratio-1} but for $T_{\rm RH}= m_s/3$. }
\label{ratio-3}
\end{figure}
For later decays, $10^9~{\rm sec} < \tau_s < t_{\rm rec}$, the photon
number preserving processes can no longer establish a Bose-Einstein
spectrum. The energy released in this case is not thermalized but 
simply heats the electrons. Through inverse-Compton scattering this
produces a distorted spectrum characterized by a  parameter $y$, which
for $|y| << 1$ is related to the energy released in non-thermal
photons as 
\begin{equation}
\frac{\Delta\rho_\gamma}{\rho_\gamma} \simeq 4y~.
\label{Deltarho2}
\end{equation}
The COBE upper bound on this parameter is $|y| < 1.5 \times
10^{-5}$~\cite{Fixsen:1996nj}. In both cases, Eqs.~(\ref{Deltarho1})
and (\ref{Deltarho2}), the fractional increase in the photon energy
density due to the decay of the sterile neutrinos can be written as
\begin{equation}
\frac{\Delta\rho_\gamma}{\rho_\gamma} \simeq f_\gamma \, B \, 
\frac{m_s \, n_s}{2.7 \, n_\gamma} \, \left(\frac{\tau_s}{\rm
  sec}\right)^{1/2} \ltap 6 \times 10^{-5}~. 
\end{equation}
All these bounds just described related to distortions in the CMB
spectrum are labeled as ``CMB'' in Figs.~\ref{ratio-1},~\ref{ratio-3} 
and~\ref{T=5}.

For decays at earlier times, the best constraints on photons produced
before the CMB thermalization epoch $t_{\rm th}$ come from BBN which
finishes by $t_{\rm endBBN} \simeq 10^4$~sec. After $t_{\rm endBBN}$,
electromagnetic cascades can cause the photodissociation of D and
${}^4{\rm He}$. For photons produced in the time interval $10^4~{\rm
sec} < \tau_s < 10^6~{\rm sec}$, the photodissociation of D poses
the best limits. For earlier and later times, the overproduction of D
due to the photodissociation of ${}^4{\rm He}$~\cite{Ellis:1990nb} far
dominates its destruction, since the abundance of ${}^4{\rm He}$ is
about 10$^4$ times greater than that of D, and sets the most stringent
constraints. The upper bounds taken from Fig. 3 of
Ref.~\cite{Ellis:1990nb} are
\begin{equation}
\left(\frac{m_s}{\rm MeV}\right) \frac{n_s}{n_\gamma} < 
10^{-2}\left(\frac{10^4 \, {\rm sec}}{\tau_s}\right)^{5/2}~,
\end{equation}
for 10$^4$sec $< \tau_s < 10^6$ sec and $m_s > 2.2~{\rm MeV}$;
\begin{equation}
\left(\frac{m_s}{\rm MeV}\right)\left(\frac{n_s}{n_\gamma}\right) <
10^{-7}\left(\frac{10^6 \, {\rm sec}}{\tau_s}\right)~,
\end{equation}
for $10^6$~sec $ < \tau_s < 10^8$~sec and $m_s > 28$~MeV and
\begin{equation}
\left(\frac{m_s}{\rm MeV}\right)\left(\frac{n_s}{n_\gamma}\right)
< 10^{-9}\left(\frac{\tau_s}{10^8 \, {\rm sec}}\right)^{1/4}~,
\end{equation}
for $10^8$~sec $< \tau_s < 10^{13}$~sec and $m_s > 28~{\rm MeV}$. We
label these bounds as ``Light Elements'' in
Figs.~\ref{ratio-1},~\ref{ratio-3} and~\ref{T=5}.

\begin{figure}[t]
\includegraphics[width=0.8\textwidth,clip=true,angle=0]{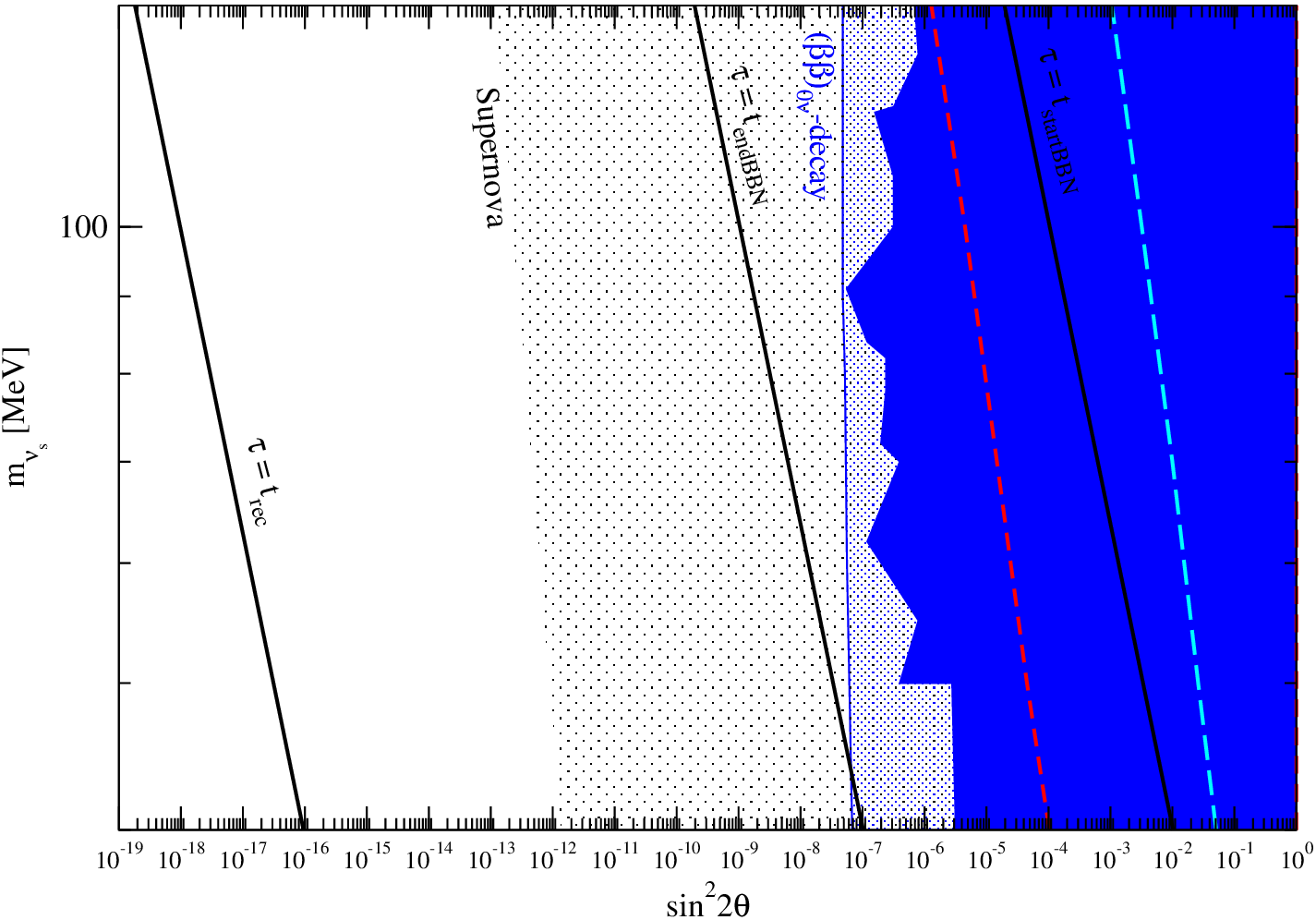} 
\caption{Same as Fig.~\ref{ratio-1} but for $T_{\rm RH} = m_s/10$.}
\label{ratio-10}
\end{figure}

For even earlier times, BBN data provides an upper bound on any source
of extra energy density present in the Universe during the BBN epoch,
$t_{\rm startBBN} \simeq 0.1$~sec $< \tau < t_{\rm endBBN}$, as well
as on extra radiation present during that period. The bounds are
complicated in detail, but it is safe to say that if the extra energy
density due to the presence of sterile neutrinos is very small, BBN
will not be affected in any way. The bounds on extra contributions to
the energy density during BBN are customarily presented in terms of the
equivalent extra number of relativistic active neutrino species $\Delta
N_\nu$. Thus, to be on the safe side, we simply require the $\Delta
N_\nu$ due to the presence of sterile neutrinos to be very small
during BBN:
\begin{equation}
\Delta N_\nu = \frac{\rho_s}{\rho_a} \simeq \frac{1}{\pi}
\frac{n_s}{n_a}\left(\frac{m_s}{\rm MeV}\right)
\left(\frac{\tau}{\rm sec}\right)^{1/2} < 1~.
\end{equation}
We label this bound as ``BBN'' in Figs.~\ref{ratio-1}, \ref{ratio-3} 
and \ref{T=5}.

Finally, there are also astrophysical disfavored regions in the
$(\sin^2 2\theta, m_s)$ space that we should mention. In principle,
the energy loss into sterile neutrinos produced in core collapse
supernovae explosions provides bounds on the mass and mixing angles of
massive (mostly sterile) neutrinos. However, so much is not understood
about the neutrino transport and flavor transformation in hot and
dense nuclear matter, that conservatively the implicated $m_s -
\sin^2{2\theta}$ region can only be considered disfavored but not
excluded. Only neutrinos with mass $m_s \ltap 150$~MeV could be
emitted copiously in the supernova core and, for those masses, mixings
$\sin^2 2 \theta_{\rm lim} \equiv 3\times10^{-10} \ltap
\sin^2{2\theta} \ltap 10^{-2}$ are
disfavored~\cite{Abazajian:2001nj,KMP91}. For small mixing angles the 
sterile neutrinos are not  trapped within the supernova, thus they are 
emitted from the whole collapsing star, mostly from its core. The
lower bound on the mixing angle is obtained by requiring that the
sterile neutrino flux at Earth emitted by the supernova 1987A,
$F_s$, is not larger than the active flux, $F_a = 1.1 \times 10^{10}
{\rm cm}^{-2}$, emitted by it. Thus, $F_s=F_a$  for  $\sin^2 2
\theta_{\rm lim}$ and for smaller angles 
\begin{equation}
F_s = \frac{\sin^2 2\theta}{\sin^2 2\theta_{\rm lim}} \, F_a.
\end{equation}
This is relevant for the the last bound we will consider. The decay of
heavy neutrinos emitted in supernova explosions would produce a
flux of photons. The non-observation by the Solar Maximum Mission 
of any $\gamma-$ray counts in excess of the background for a time
interval of $t_{\rm max} =$~223.2~sec after the arrival of the first
$\overline{\nu}_e$'s from supernova 1987A, allows us to enlarge the 
disfavored regions of the parameter space. Following the analysis of
Ref.~\cite{SNrad} and neglecting the absorption of the photons produced
 in decays within the supernova, the $\gamma$ flux is given by
\begin{equation}
\frac{d F_\gamma}{d E_\gamma} = 
\frac{F_s \, E_{\gamma} \, f_{\gamma} \, B}{T \, m_s \, \tau} 
\, \int_{0}^{t_{\rm max}} \, dt \, e^{-2 \, E_{\gamma} \, t / m_s
  \, \tau} \, \left( 1 + \frac{E_{\gamma}}{T} \right)  \,
e^{-E_{\gamma}/T}~.
\label{Snradfluxeq}
\end{equation}
Here $T \simeq 50$~MeV is the temperature at which sterile neutrinos
are emitted from the supernova core.   We use the  3-$\sigma$ limits
$(\Delta F_{\gamma})_{3-\sigma} (E_i,E_f)$ on the $\gamma$ flux for
the time interval considered obtained in Ref.~\cite{SNrad} for three
energy bands $(E_i,E_f)$, namely $(4.1, 4.4)$~MeV, $(10,25)$~MeV and
$(25,100)$~MeV (see Table 1 of Ref.~\cite{SNrad}). We calculate the
disfavored region in the $(\sin^2 2\theta, m_s)$ parameter space  by
requiring
\begin{equation}
\int_{E_i}^{E_f} \, \frac{dF_{\gamma}}{dE_{\gamma}} \, dE_{\gamma} <
(\Delta F_{\gamma})_{3-\sigma} (E_i,E_f)~.
\end{equation}
In the limit $m_s \, \tau >> 2 \, t_{\rm max} \, T$, which holds in
the region of interest, this condition can be written as 
\begin{equation}
\frac{F_s \, f_{\gamma} \, B}{2} \, \frac{2 \, t_{\rm max} \, T}{m_s
  \, \tau} \, G(E_i,E_f) < (\Delta F_{\gamma})_{3-\sigma} (E_i , E_f)~,
\end{equation}
with 
\begin{equation}
G(E_i,E_f) \equiv 3 \, \left(e^{-E_i/T} - e^{-E_f/T}\right) + 3 \,
\left(\left(\frac{E_i}{T}\right) \, e^{-E_i/T} -
\left(\frac{E_f}{T}\right) \, e^{-E_f/T}\right) + 
\left(\left(\frac{E_i}{T}\right)^2 \, e^{-E_i/T} -
\left(\frac{E_f}{T}\right)^2 \, e^{-E_f/T}\right)~.
\end{equation}
The band which gives the most restrictive limits for the range of
sterile neutrino masses considered is $(25,100)$~MeV, although the
band $(10,25)$~MeV gives comparable results. The two types of regions
disfavored by core collapse supernovae arguments are the the dotted
regions  labeled  ``Supernova'' in
Figs.~\ref{ratio-1},~\ref{ratio-3},~\ref{ratio-10} and~\ref{T=5}.

\begin{figure}[t]
\includegraphics[width=0.8\textwidth,clip=true,angle=0]{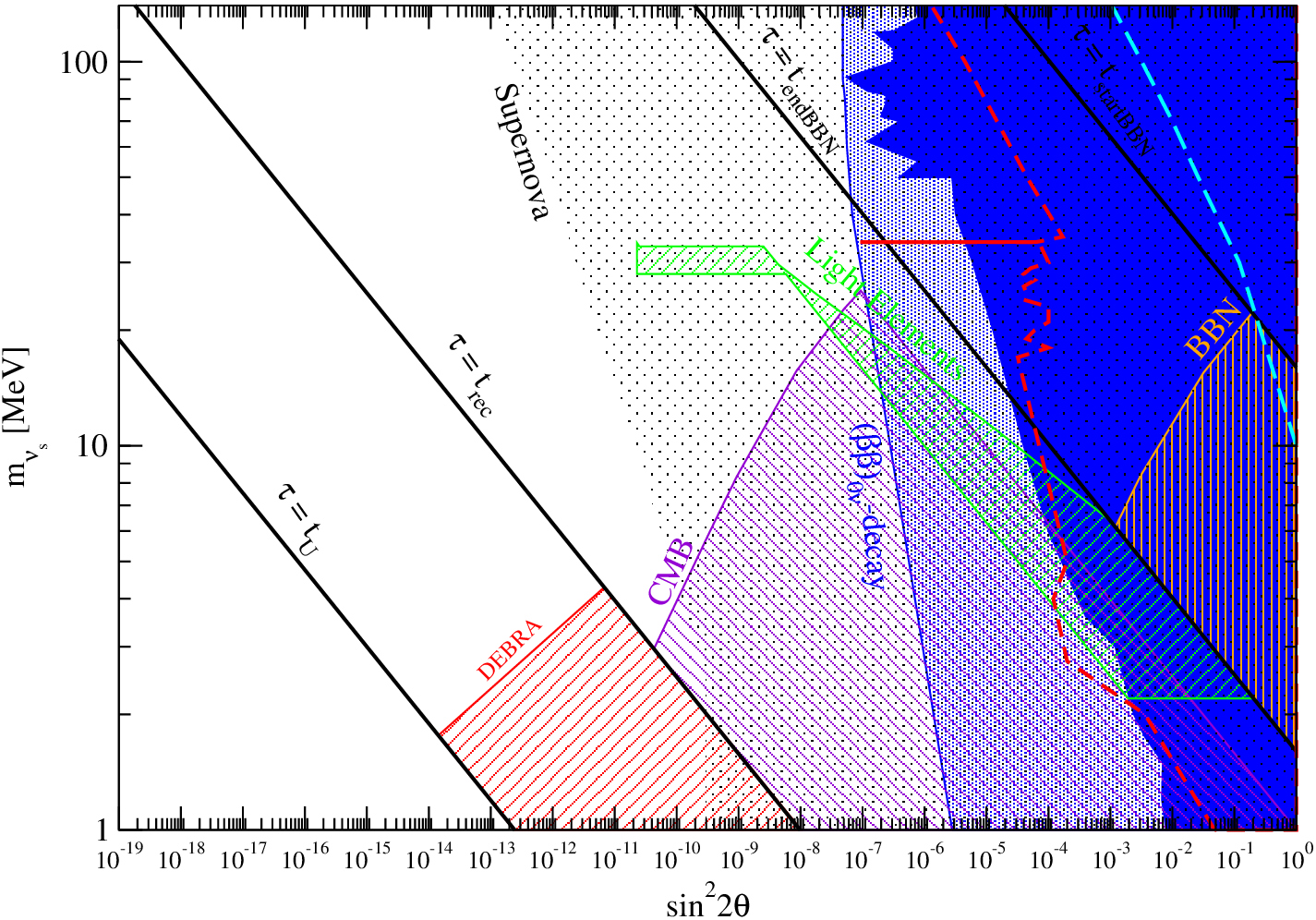} 
\caption{Same as Fig.~\ref{ratio-1} but for a fixed value of the
  reheating temperature, $T_{\rm RH}= 5$~MeV and for heavy neutrino
  masses higher than 1~MeV.}  
\label{T=5}
\end{figure}

Figs.~\ref{ratio-1} to~\ref{T=5} show in the $(\sin^2 2 \theta, m_s)$
parameter space all the bounds on active-sterile neutrino mixing we
have presented above, for $T_{\rm RH} > 4$~MeV and 1~MeV~$< m_s <
140$~MeV. For concreteness, in all the figures we have taken $f_\gamma
= 1$ and $B=0.1$. Each cosmological bound is indicated by its
corresponding label, as explained above. The regions globally excluded
by laboratory measurements (Section~\ref{lab}) are: the dark gray
(blue) area for $\nu_e-\nu_s$ mixing, the region to the right of the
red short-dashed line for $\nu_\mu-\nu_s$ mixing and that to the right
of the (magenta) long-dashed line for $\nu_\tau-\nu_s$ mixing. In the
case of $\nu_e-\nu_s$ mixing the bounds for Majorana neutrinos are
more restrictive than for Dirac neutrinos (see the hatched dark gray
(blue) area labeled as ``$(\beta \beta)_{0 \nu}-$decay''). Values 
of the heavy neutrino lifetime $\tau$ equal to the relevant epochs in
the history of the Universe are also shown. 

In Figs.~\ref{ratio-1}, \ref{ratio-3} and \ref{ratio-10}, we display
the results for three different values of the ratio $x_{\rm RH} \equiv
m_s/T_{\rm RH}$, $x_{\rm RH} = 1, 3, 10$, respectively. As expected,
for $T_{\rm RH}$ increasingly smaller than the sterile neutrino mass,
the cosmological bounds become less restrictive, and when $T_{\rm RH}
\leq m_s/10$, the cosmological bounds become completely irrelevant and
only experimental data are able to restrict the parameter space. This
result can also be seen in Fig.~\ref{T=5}, in which the reheating
temperature is fixed to be $T_{\rm RH} =5$~MeV. In this case all
cosmological bounds become irrelevant for $m_s \geq 30$~MeV.

\section{Conclusions}
\label{conclusions}

Sterile neutrinos are invoked in many extensions of the Standard Model
of particle physics~\cite{MeVDM,nuMSM,degouvea,ETC}. However, it is
commonly assumed that the cosmological and astrophysical bounds on the
mixings of sterile and active neutrinos restrict the range of their
allowed values much more than laboratory data. In fact, sterile
neutrinos with parameters suitable to be found in the near future in
different experiments would have mixings with active neutrinos too
large to be allowed by the standard cosmological assumptions about the
pre-BBN era in the Universe, an era about which we do not have any
observational information. The standard assumptions are few but very
powerful: it is usually assumed that the temperature reached in the
radiation dominated epoch before BBN was very high, that the Universe
was radiation dominated then and that the entropy of radiation and
matter is conserved.

Here, we show that it is possible to evade most of the cosmological
bounds by assuming that the temperature at the end of (the last
episode of) inflation or entropy production, the so-called reheating
temperature $T_{\rm RH}$, is low enough. We concentrate on massive
(mostly sterile) neutrinos heavier than 1~MeV, having previously dealt
with the lighter ones~\cite{Gelmini:2004ah}. For low $T_{\rm RH}$, the
production of sterile neutrinos is suppressed as shown in
Fig.~\ref{ns}. For example, going from $T_{\rm RH} = m_s$ to $T_{\rm
  RH} = m_s/10$ there is a suppression of four orders of magnitude. We
present the experimental bounds on sterile neutrino mixings and  the
cosmological bounds imposed by the diffuse extragalactic background
radiation, the CMB, BBN and the abundance of light elements. We find
that for $T_{\rm RH} $ a few times smaller $m_s$ the cosmological
bounds weaken significantly and for $T_{\rm RH} = m_s/10$ they
disappear completely. This  is shown in
Figs.~\ref{ratio-1},~\ref{ratio-3},~\ref{ratio-10}.  In Fig.~\ref{T=5}
we keep instead the reheating temperature at a fixed value $T_{\rm RH}
= 5$~MeV, close to the lower bound on $T_{\rm RH}$ of 4~MeV imposed by
BBN and other cosmological data. In this case, no cosmological bounds
remain for $m_s > 30$~MeV, thus the only constraints on sterile
neutrinos in this mass range come from terrestrial experiments. Hence,
unlike in the standard cosmology, in low reheating temperature
cosmologies it is possible to accommodate ``visible" sterile
neutrinos, i.e., sterile neutrinos which could soon be found in
experiments.

Cosmological scenarios with a very low reheating temperature are more
complicated than the standard one. Although no consistent
all-encompassing model of this nature exists at present, different
aspects have been studied with interesting results, which suggest that
a coherent scenario could be produced if an experimental indication
would lead us to it. In fact, finding a particle, such as a ``visible"
sterile neutrino, whose existence would contradict the usual
assumptions about the pre-BBN era, would give us not only invaluable
information for particle physics, but also an indication of enormous
relevance in cosmology: it would tell us that the usual assumptions
must be modified, for example in the manner presented in this paper.

\section*{Acknowledgments}
SPR and SP thank UCLA for hospitality at the initial stages of this
work. SP also thanks CERN for hospitality. GG was supported in part by
NASA grants US DOE grant DE-FG03-91ER40662 Task C, and NASA grants
NAG5-13399 and ATP03-0000-0057. SPR is partially supported by the
Spanish Grant FPA2005-01678 of the MCT.

\section*{Appendix} 

To be formally accurate, the derivation of Eq.~(\ref{Boltzmann-2})
should be carried out in the mass basis, since  the distribution
function is defined for species of definite energy. In the vacuum 
limit, in which we are interested, one can start with the Boltzmann
equation in terms of the 2 by 2 density matrix $\rho$ acting on the
mass eigenstates $\nu_1$ and $\nu_2$, with elements $\rho_{11},
\rho_{12}, \rho_{21}, \rho_{22}$~\cite{dolgov3}, 
\begin{equation}
i\dot{\rho}=-[H,\rho]-i\{\Gamma,(\rho-\rho_{eq})\}
\label{density matrix}~,
\end{equation}
where $\rho_{eq}$ is given in terms of the equilibrium momentum
distributions $f_i^{\rm eq}$ for $i=1,2$, $ f_i^{\rm eq} =
(e^{(E_i-\mu_i)/T}+1)^{-1}$ ($E_i$ and $\mu_i$ are the energy and
chemical potential of the $i$ mass eigenstate and without a lepton
asymmetry $\mu_i=0$), 
\begin{equation}
\rho_{eq}= \left[ \begin{array}{ c c }  f_1^{\rm eq} & 0 \\
     0 & f_2^{\rm eq} \end{array} \right],
\label{distribution}
\end{equation}
and the 2 by 2 matrix $ H$ is
\begin{equation}
H= \left[\begin{array}{c c} E_1 &0 \\ 0 & E_2 \end{array}\right]~.  
\label{H}
\end{equation}
Neutrino production and destruction are represented by the
anticommutator term  in Eq.~(\ref{density matrix}), which describes 
the coherence-breaking interactions. The matrix $\Gamma$ can be
written as
\begin{equation}
 \Gamma = \left[\begin{array}{c c} \cos^2{\theta} (\Gamma/2+\delta
 \Gamma_1)&  -\cos{\theta}\sin{\theta}(\Gamma/2+\delta \Gamma) \\ 
  -\cos{\theta}\sin{\theta} (\Gamma/2+\delta \Gamma)&
 ~\sin^2{\theta}(\Gamma/2 +\delta \Gamma_2) \end{array}\right],
  \label{gamma description}
\end{equation}
where $\Gamma$ is the interaction rate for massless neutrinos, which
is of order $G_{\rm F}^2$ (i.e., second order in the Fermi coupling
constant), $\delta\Gamma_i$ are  the corrections necessary when the
incoming and outgoing particles have a non-zero mass $m_i$ in the case
of scattering, or when the two annihilating particles are of mass
$m_i$ and $\delta\Gamma$ is the correction necessary when one of the
two interacting particles has mass $m_1$ and the other has mass
$m_2$. 

Replacing Eqs.~(\ref{distribution}), (\ref{H}) and (\ref{gamma
description}) in Eq.~(\ref{density matrix}) we can find the equation
for $\dot{\rho_{11}}$, $\dot{\rho_{22}}$, and $\dot{\rho_{12}}$, 
which is the complex conjugate of $\dot{\rho_{21}}$. Writing $i
\dot{\rho_{12}}=R+ iI$ and $i \dot{\rho_{21}}=R- iI$, one can find the
equations for the time derivative of the real and imaginary  parts
$\dot R$ and $\dot I$ in terms of $R$ and $I$. As demonstrated in
Ref.~\cite{BVW99}, in the stationary point or static approximation,
we can take $\dot R = \dot I = 0$ which allows us to solve for $R$
and $I$. Using the expression for $R$ and $I$ so obtained and assuming
that the less massive eigenstate $\nu_1$ maintains equilibrium, for
small values of $\sin^2{2\theta}$ one gets an expression for
$\dot{\rho_{22}}$ which coincides with Eq.~(\ref{Boltzmann-2}) in
vacuum (once we take $\rho_{22} = f_2 \simeq f_s$ and $f_2^{\rm eq}
\simeq f_s^{\rm eq}$).

\end{document}